# Tunable magnetic exchange interactions in manganese-doped inverted core/shell ZnSe/CdSe nanocrystals

David A. Bussian[1,2], Scott A. Crooker[3*], Ming Yin[1], Marcin Brynda[4], Alexander L. Efros[5] & Victor I. Klimov[1,2*]

[1]Chemistry Division, Los Alamos National Laboratory, Los Alamos, New Mexico 87545

[2]Center for Integrated Nanotechnologies, Los Alamos, New Mexico 87545

[3]National High Magnetic Field Laboratory, Los Alamos, New Mexico 87545

[4]Department of Chemistry, University of California, Davis, California 95616

[5]Naval Research Laboratory, Washington, DC 20375

*e-mail: crooker@lanl.gov, klimov@lanl.gov

**Magnetic doping of semiconductor nanostructures is actively pursued for applications in magnetic memory and spin-based electronics[1,2]. Central to these efforts is a drive to control the interaction strength between carriers (electrons and holes) and the embedded magnetic atoms[3-5]. In this respect, colloidal nanocrystal heterostructures provide great flexibility via growth-controlled 'engineering' of electron and hole wavefunctions within individual nanocrystals[6,7]. Here we demonstrate a widely tunable magnetic *sp-d* exchange interaction between electron-hole excitations (excitons) and paramagnetic manganese ions using 'inverted' core-shell nanocrystals composed of $Mn^{2+}$-doped ZnSe cores overcoated with undoped shells of narrower-gap CdSe. Magnetic circular dichroism studies reveal giant Zeeman spin splittings of the band-edge exciton that, surprisingly, are tunable in both magnitude *and sign*. Effective exciton g-factors are controllably tuned from -200 to +30 solely by increasing the CdSe shell thickness, demonstrating that strong quantum confinement and wavefunction engineering in heterostructured nanocrystal materials can be utilized to manipulate carrier-$Mn^{2+}$ wavefunction overlap *and* the *sp-d* exchange parameters themselves.**

Traditionally, embedding paramagnetic atoms into low-dimensional semiconductor structures requires molecular-beam epitaxy or chemical vapor deposition techniques [3-5]. There now exists a rich variety of 'diluted magnetic semiconductor' (DMS) quantum wells, superlattices, and hetero-interfaces, with recent work demonstrating magnetic doping of epitaxially-grown 'zero-dimensional' quantum dots [8-10]. In parallel however, advances in colloidal chemistry have recently allowed magnetic doping of semiconductor nanocrystals (NCs)[11-16], providing an alternative and potentially lower-cost route towards magnetically active quantum dots. With a view towards enhancing carrier-paramagnetic ion spin interactions, colloidal NCs typically generate stronger spatial confinement of electronic wavefunctions compared to their epitaxial counterparts, which is thought to enhance $sp$-$d$ exchange coupling even for a single magnetic dopant atom[12,17].

While magnetically-doped monocomponent NCs are well established[16], wavefunction engineering using magnetic multi-component colloidal *heterostructures*[18,19] has not been extensively explored. One new class of NC heterostructure that holds promise for tuning $sp$-$d$ exchange interactions are 'inverted' core-shell designs, wherein wide-gap semiconductor cores are overcoated with narrower-gap shells. With increasing shell thickness, the electron and hole envelope wavefunctions, $\Psi_{e,h}(r)$, migrate towards the NC periphery[7] (albeit at different rates), thus tuning their overlap with magnetic atoms embedded, for example, in the core alone. In this work, we investigate precisely this type of wavefunction engineering and exchange interaction control using 'inverted' ZnSe/CdSe core/shell NCs whose cores are doped with paramagnetic, spin-$\frac{5}{2}$ $Mn^{2+}$ ions (see Fig. 1). Magnetic circular dichroism (MCD) spectroscopy at the NC absorption edge reveals a giant $sp$-$d$ exchange interaction that *inverts sign* with increasing shell thickness, suggesting a confinement-induced sign inversion of the electron-$Mn^{2+}$ exchange constant, $\alpha$, accompanied by significant reduction of the hole-$Mn^{2+}$ overlap due to wavefunction engineering.

Four series of ZnSe/CdSe NCs were grown, each having ZnSe cores of radius $r \cong 17$ Å. Within each series the CdSe shell thickness, $h$, systematically increases from 0-8 Å. Two series used nonmagnetic (undoped) 'reference' cores, and two used $Mn^{2+}$-doped cores. Elemental analysis of pyridine-washed magnetic cores indicates ~2 $Mn^{2+}$ ions per core, on average. Paramagnetic resonance studies are consistent with the $Mn^{2+}$ residing primarily within the ZnSe core, for all $h$ (Supplementary Information). Figure 1a shows absorption and photoluminescence (PL) spectra from the $Mn^{2+}$:ZnSe cores alone ($h$=0). The absorption peak at ~3.2 eV is due to the fundamental band-edge (1S) exciton transition. On the other hand, the PL is dominated by the 2.15 eV internal $^4T_1 \rightarrow {}^6A_1$ $Mn^{2+}$ transition, which results from efficient energy transfer from band-edge excitons to the excited $^4T_1$ Mn$^{2+}$ state[11]. The PL also shows a small peak at 2.95 eV from direct recombination of band-edge excitons.[14,19]

To suppress energy transfer and realize strong exciton PL, the NC bandgap must be tuned below ~2.15 eV, as recently demonstrated in Mn$^{2+}$:CdSe NCs[20]. This regime is accessible here using sufficiently thick shells (see Fig. 1b). Increasing $h$ shifts $\Psi_e(r)$ and $\Psi_h(r)$ towards the shell, reducing the bandgap and red-shifting both the absorption and the PL. When $h \geq 5$ Å, the exciton PL energy drops below 2.15 eV. The dependence of the exciton PL energy on the absorption energy is summarized in Fig. 1c, for both Mn$^{2+}$-doped and nonmagnetic NCs.

The most compelling evidence for successful Mn$^{2+}$ incorporation is an enhanced exciton Zeeman splitting, $\Delta E_Z$, due to carrier-Mn$^{2+}$ *sp-d* exchange. Low-temperature MCD spectroscopy provides a direct, quantitative measure of $\Delta E_Z$ at the fundamental 1S absorption peak[21]. Figure 2 shows $\Delta E_Z$ versus magnetic field ***H*** at different temperatures for four important cases: nonmagnetic and Mn$^{2+}$-doped ZnSe/CdSe NCs, each having 'thin' and 'thick' shells. Nonmagnetic NCs with thin or thick shells show derivative-like MCD signals at the absorption peak that increase linearly with field from 1-6 T (Figs. 2a,b, insets). The exciton Zeeman splittings, $\Delta E_Z = g_{ex}\mu_B H$, indicate small, temperature-independent, positive exciton g-factors of order unity ($g_{ex}$=+2.1 and +2.5, respectively), in approximate agreement with previous studies of nonmagnetic monocomponent NCs[14,21]. As expected, shell thickness does not strongly influence the small intrinsic electron and hole g-factors in these nonmagnetic NCs.

*Magnetic* NCs having thin shells (Fig. 2c) exhibit much larger MCD of opposite sign. $\Delta E_Z$ is negative, quite large, and tracks the temperature-dependent magnetization of the paramagnetic Mn$^{2+}$ ions, which is described by a Brillouin function (dashed line). These data reveal a strong *sp-d* exchange interaction between the absorption-edge exciton and the Mn$^{2+}$ in the ZnSe core, and are qualitatively similar to previous studies[12,14,15] of monocomponent ZnSe or CdSe NCs doped with Mn$^{2+}$ or Co$^{2+}$.

In surprising contrast, Fig. 2d reveals that in Mn$^{2+}$-doped NCs with *thick* CdSe shells, $\Delta E_Z$ has *opposite* sign, yet still exhibits a significant *sp-d* coupling: $\Delta E_Z$ still saturates at low temperatures and high fields and is significantly larger than in nonmagnetic NCs. Carrier-$Mn^{2+}$ *sp-d* exchange coupling is clearly still appreciable; however, its strength has been *tuned through zero*. A clear inversion of the *sp-d* exchange is seen in Fig. 3a, which shows $\Delta E_Z$ versus $H$ at 1.6 K for one series of $Mn^{2+}$-doped ZnSe/CdSe NCs. Within this series, the 1S absorption edge drops from 2.9 eV (red trace) to 2.2 eV (purple trace) as $h$ increases to 8Å. For *all* NCs within this magnetic series, $\Delta E_Z$ exhibits saturation at high magnetic fields (and also reveals a Brillouin-function temperature dependence; not shown) indicating significant *sp-d* interactions. $\Delta E_Z$ inverts when $h$~2Å. For comparison the small, linear $\Delta E_Z$ from nonmagnetic NCs is also shown (black trace). Effective exciton *g*-factors derived from the low-field Zeeman splitting, $g_{eff}$, along with $g_{eff}$ from all other NC series, are plotted in Fig. 3b versus the absorption energy. With increasing $h$, $g_{eff}$ is tuned from -200 to +30.

While elegant experiments in DMS quantum wells have shown that the net *sp-d* interaction's *magnitude* is tunable via wavefunction engineering[22], to our knowledge its *sign* has never been shown to invert, which is unexpected within the traditional framework of *sp-d* exchange (discussed below) and which points to new physics in these core-shell NC materials. In the following we show that the inversion of $\Delta E_Z$ can arise, in part, from the strong 'zero-dimensional' quantum confinement realized in NCs, which exceeds that typically found in 2-D heterostructures, and which can invert the sign of $\alpha$, the electron-$Mn^{2+}$ exchange parameter.

Although a single $Mn^{2+}$ spin interacting with a single exciton will generate - even in zero field - a measurable $\Delta E_Z$ in a single quantum dot[10,12], our colloidal NC cores each contain ~2 $Mn^{2+}$ spins on average, whose overlap-weighted average magnetization determines $\Delta E_Z$ in any given NC. Further, the $Mn^{2+}$ in the ZnSe cores are positioned randomly, and the number of $Mn^{2+}$ per core fluctuates statistically, allowing us to discuss *sp-d* exchange interactions in these NC ensembles in terms of a pseudo-bulk model. Here, spin splittings of the conduction and valence band edges in the NCs have two contributions; namely 1) a small splitting from the 'intrinsic' electron and hole *g*-factors in the host semiconductor, $g_{e,h}$ (of order unity in CdSe and ZnSe), and 2) a potentially much larger splitting from the *s-d* (*p-d*) exchange interaction between the $s_e=\frac{1}{2}$, *s*-like electrons ($j_h=\frac{3}{2}$, *p*-like holes) and the $S=\frac{5}{2}$ *d*-shell moments of the $Mn^{2+}$. The *s-d* and *p-d* character of the electron-$Mn^{2+}$ and hole-$Mn^{2+}$ interaction are characterized by the exchange energies $N_0\alpha$ and $N_0\beta$, respectively, where $N_0$ is the density of unit cells. In bulk II-VI DMS[3,23], $N_0\alpha$ arises from potential (ferromagnetic) s-d exchange and is positive, while $N_0\beta$ derives predominantly from kinetic-type (antiferromagnetic) p-d exchange and is larger and negative. In bulk $Zn_{1-x}Mn_xSe$ ($Cd_{1-x}Mn_xSe$), $N_0\alpha$ = +0.29 eV (+0.23 eV) and $N_0\beta$ = -1.4 eV (-1.27 eV). [23]

In analogy with bulk DMS materials[3,23], s-d and p-d exchange in our core/shell NCs generates a spin splitting between the $s_z = \pm\frac{1}{2}$ electrons and the $j_z = \pm\frac{3}{2}$ holes equal to $f_e N_0\alpha <S_z>$ and $f_h N_0\beta <S_z>$, respectively. In our NCs, $f_e$ and $f_h$ characterize the degree of spatial overlap between the $Mn^{2+}$ ions in the core and the modulus-square of the wavefunctions $\Psi_e(r)$ and $\Psi_h(r)$ (in bulk DMS, $f_e$ and $f_h$ simply equal the average $Mn^{2+}$ concentration). In an ensemble of NCs, $f_e$ and $f_h$ can therefore be regarded as the average probability that the electron and hole reside in the core, $\langle |\Psi_{e,h}(r)|^2 \rangle$, multiplied by the average number of $Mn^{2+}$ per core. The last quantity, $<S_z>$, is the average spin projection per $Mn^{2+}$ along $H$. For low $Mn^{2+}$ doping, $<S_z>$ follows a Brillouin function, which describes the magnetization of paramagnetic $Mn^{2+}$ moments; *i.e.*, $<S_z>$ saturates at low temperatures and high magnetic fields. By convention $<S_z>$ is negative, being oriented antiparallel to $H$.

In colloidal II-VI NCs, experiments and theory have established[24] that band-edge, 1S excitons (nominally composed of twofold degenerate $s_e$=½ electrons and fourfold degenerate $j_h$=3/2 holes) are split by the effects of strong electron-hole exchange, crystal field, and shape asymmetry into five distinct levels labeled by $F$, the total angular momentum projection on the NC symmetry axis: $F = 2, 1^L, 0^L, 1^U$, and $0^U$, where "U/L" denotes upper/lower levels. Figure 4a shows the ordering of these levels in nearly-spherical NCs. Exciton PL originates primarily from the lower optically-active state ($1^L$), whereas the band-edge absorption peak derives from the upper $1^U$ transition, which has much larger oscillator strength. The energy separating the $1^{L,U}$ states gives the large 'global' Stokes shift typically observed in NCs (see Fig. 1). Both $1^U$ and $1^L$ excitons are two-fold degenerate with respect to total angular momentum ($F=\pm 1^{L,U}$). Thus, the MCD originates in the Zeeman spin splitting of the absorbing $\pm 1^U$ states, which couple to $\sigma^\pm$ photons respectively.

In our ZnSe cores, the $1^U$ transition is composed largely (75%) of transitions from $j_z = \pm\frac{3}{2}$ hole states, with a small (25%) admixture from $j_z = \pm\frac{1}{2}$ hole transitions[24]. Neglecting the latter contribution and using the standard selection rules for the absorption of circularly-polarized light, we can express $\Delta E_Z$ measured in these NC ensembles as:

$$\Delta E_Z=(3g_h-g_e)\mu_B H+\langle S_z\rangle N_0(f_e\alpha-f_h\beta), \quad \text{(Eq. 1)}$$

which is very similar to that in conventional DMS materials. Note that in bulk DMS, where $\alpha>0$ and $\beta \cong -4\alpha$, the small 'intrinsic' contribution to $\Delta E_Z$ (first term; of order +0.1 meV/T in ZnSe) is typically overwhelmed by the large and negative *sp-d* contribution, which can exceed -100 meV at low temperatures.

Equation 1 provides a good description of the average $\Delta E_Z$ in these ensembles of $Mn^{2+}$-doped NCs, and hence the observed *inversion* of $\Delta E_Z$ indicates a sign reversal of the average exchange term, $(f_e\alpha-f_h\beta)$. Clearly, simply modifying the overlap integrals $f_{e,h}$ by wavefunction engineering cannot invert this quantity if $\alpha$ and $\beta$ retain the same sign as in the bulk (positive and negative, respectively). Rather, the sign of $\alpha$ or $\beta$ in these NCs *must be different* than in the bulk. Such a possibility is indeed expected based on recent experiments in DMS quantum wells[25-27] and by theory[26,28] which established that $\alpha$ changes with quantum confinement and may even invert (become negative) in strongly-confined II-VI quantum dots (although this has never yet been observed). This effect results from an admixture of *p*-type valence band symmetry into the electron's Bloch wavefunction, causing a negative kinetic-exchange contribution to $\alpha$ that increases with confinement energy ($\beta$ remains largely unaffected, being already dominated by kinetic exchange).

Following Merkulov[26], in quantum-confined systems $\alpha$ can be approximated as $\alpha = \alpha_{bulk}+|C_v|^2\gamma(E)\beta_{bulk}$, where the coefficient $|C_v|^2 \sim \Delta E_e/E_g$ describes the valence-band contribution to the electron's Bloch wavefunction, which scales as the ratio of the electron confinement energy, $\Delta E_e$, to the bulk bandgap $E_g$. The kinetic exchange parameter $\gamma(E)$ is calculated in second-order perturbation theory and depends strongly (resonantly) on the proximity of the confined electron and hole energies ($E_e$, $E_h$) to the occupied and unoccupied levels of the $Mn^{2+}$ 3d electrons ($\varepsilon^+,\varepsilon^-$):

$\gamma(E) = (E_h - \varepsilon^+)(\varepsilon^- - E_h)/(E_e - \varepsilon^+)(\varepsilon^- - E_e)$ (see Fig. 4c). Strong confinement in ZnSe NCs influences $\gamma(E)$ primarily through the resonant $(\bar{\varepsilon}-E_e)^{-1}$ term, since $E_e$ shifts much closer to $\bar{\varepsilon}$ than in the bulk. Using literature values[29] ($\varepsilon^+$ is ~3.5 eV below the valence band edge and $\bar{\varepsilon}$ is ~3.5-4.0 eV above it), we estimate that $\alpha$ inverts sign (becomes negative) when $\Delta E_e$>200-300 meV. In our 17Å radius ZnSe cores, the 1S absorption edge blueshifts by ~400 meV (see Fig. 1), indicating that $N_0\alpha$ is likely *negative* and of order -0.2 eV.

However, without a significant disparity between the carrier-$Mn^{2+}$ overlap integrals $f_{e,h}$, a small negative value of $\alpha$ alone will not invert $\Delta E_Z$. In NCs with thin CdSe shells, $\Psi_e(r)$ and $\Psi_h(r)$ reside primarily in the core, so that $f_e \approx f_h$ and $\Delta E_Z$ remains dominated by the large hole-$Mn^{2+}$ coupling and is negative, as experimentally observed (Fig 2c). Inverting $\Delta E_z$ requires a marked and preferential reduction of hole-$Mn^{2+}$ overlap, such that $f_h << f_e$. Precisely this situation occurs in NCs with thicker CdSe shells ($h > \sim 2$Å), where $\Psi_h(r)$ migrates to the nonmagnetic shell more rapidly than $\Psi_e(r)$ (see Fig. 1), effectively 'turning off' the hole-$Mn^{2+}$ coupling. Though not expected to occur within a model of S-like wavefunctions only[7], this disparity between $f_e$ and $f_h$ arises because, unlike $\Psi_e(r)$, $\Psi_h(r)$ has both S- *and* D-like spatial symmetry[24] (*i.e.*, its radial wavefunction has both $j_0$ and $j_2$ spherical Bessel components; the latter vanishes at the NC center and concentrates near the nanocrystal surface even in core-only NCs). The relative contribution from $j_2$ increases with $h$ because thicker shells favor localization of holes with D-type symmetry. Thus, $f_h$ rapidly decreases to zero with increasing shell thickness, suppressing *p-d* exchange and inverting the *sp-d* exchange, $N_0(f_e\alpha - f_h\beta)$, as experimentally measured. Figures 4a,b explicitly diagram how the relevant energy levels evolve with increasing $h$, in both an exciton ($\pm 1^U$) and in a separate electron-hole ($s_e$, $j_h$) picture (see figure caption for details). As demonstrated here, the extremely strong 'zero-dimensional' quantum confinement afforded in heterostructured NC leads to new regimes of tunable carrier-$Mn^{2+}$ spin exchange in these materials, and future measurements are aimed at resolving, separately, the electron and hole exchange parameters.

# METHODS

**CORE/SHELL NANOCRYSTAL SYNTHESIS AND CHARACTERIZATION**
Growth of the $Mn^{2+}$-doped, $r \cong 17$ Å ZnSe cores followed the procedure described in Ref. 14. Overcoating with CdSe followed Ref. 7, where 5 mL of ZnSe NCs in hexadecylamine were transferred to 8 mL of dry trioctylphosphine oxide (TOPO) at 140 °C under nitrogen flow. Then, a mixture of 4 mL TOP, 0.25 mL 2M TOPSe, and 30 μL dimethylcadmium ($CdMe_2$) was slowly injected into the reaction. The temperature was gradually increased to 200 °C over 3 - 4 days while 4-5 additional injections of the TOP, TOPSe, and $CdMe_2$ mixture were conducted. Photoluminescence (PL) and absorption spectroscopy were used to monitor the growth of the CdSe shell. To quench the reaction, the temperature was lowered to 100 °C and the sample was quickly mixed with hexane. Core and (thick) shell dimensions were also directly measured by transmission electron microscopy. The $Mn^{2+}$ concentration from elemental analysis of pyridine-washed $Mn^{2+}$-doped ZnSe cores was determined by ICP-OES (inductively coupled plasma – optical emission spectroscopy) to be ~0.4% ± 0.1% of all cations, corresponding to ~2 $Mn^{2+}$ ions per ZnSe core, on average.

**MAGNETIC CIRCULAR DICHROISM MEASUREMENTS**

MCD measures the normalized difference between the transmission of right- and left-circularly polarized light through the NC sample in the Faraday geometry, $(T_R-T_L)/(T_R+T_L)$, as a function of photon energy. For $\Delta E_Z$ small compared to the Gaussian width, $\sigma$, of the fundamental 1S absorption peak, the MCD spectrum is derivative-like with a low-energy maximum amplitude $I_{max}$ that is proportional to the Zeeman splitting: $\Delta E_Z = -2\sigma I_{max}/A_{max}$, where $A_{max}$ is the absorbance of the NCs at $I_{max}$.[21] Low temperature MCD studies were performed on thin films of core/shell NCs mounted in the variable temperature insert (1.5-300 K) of an 8T superconducting magnet with direct optical access. Spectrally narrow (<0.5 nm) probe light of tunable wavelength was derived from a Xe lamp directed through a 0.3 m scanning spectrometer. Before being focused through the NC film, the probe beam was mechanically chopped at 137 Hz and its polarization was modulated between right- and left-circular at 84 kHz using a linear polarizer and a photoelastic modulator (PEM). A silicon avalanche photodiode detected the light transmitted through the sample, and $T_R-T_L$ and $T_R+T_L$ were extracted using lock-in amplifiers referenced to the PEM and to the chopper, respectively.

**FITTING TO A BRILLOUIN FUNCTION**

In the $Mn^{2+}$-doped NCs, $\Delta E_Z$ is observed to saturate at low temperatures and high magnetic fields, tracking the average magnetization per $Mn^{2+}$ ion, $<S_z>$, which is defined as a negative quantity (antiparallel to $H$). To account for antiferromagnetic correlations and clustering among neighboring $Mn^{2+}$ ions, $<S_z>$ is typically described by a modified Brillouin function, $\langle S_z \rangle = S_z^{sat} B_S \left[ \frac{g_{Mn} \mu_B S H}{k_B (T+T_0)} \right]$, where $T + T_0$ is an effective temperature, and $B_S(x) = \frac{2S+1}{2S} \coth\left[\frac{2S+1}{2S} x\right] - \frac{1}{2S} \coth\left[\frac{1}{2S} x\right]$ is the Brillouin function. $Mn^{2+}$ ions have total spin $S$=5/2. $S_z^{sat}$ is the effective saturation value of $<S_z>$ per $Mn^{2+}$ ion, which may be considerably smaller than -5/2 when the average $Mn^{2+}$ concentration $x_{Mn}$>~1%. At the low $Mn^{2+}$ doping levels in these core/shell NCs, the $Mn^{2+}$ ions are largely isolated paramagnets. However, some Mn-Mn correlations are revealed by the nonzero fitting values of $T_0$ (e.g., in Fig. 2, where the increase in T+$T_0$ from 4K to 9K suggests additional Mn diffusion -- and increased Mn-Mn interactions -- during the 3-4 days required to grow thick CdSe shells).

**Acknowledgements**

We thank B. Prall for technical assistance. This work was supported by Los Alamos LDRD Funds and the Chemical Sciences, Biosciences, and Geosciences Division of the Office of Basic Energy Sciences, Office of Science, U.S. Department of Energy (DOE). D.A.B. and V.I.K. are partially supported by the DOE Center for Integrated Nanotechnologies jointly operated by Los Alamos and Sandia National Laboratories.  A.L.E. acknowledges financial support from ONR.

**FIGURES**

**Figure 1. An 'inverted core-shell' approach to tuning *sp-d* spin-exchange interactions in heterostructured colloidal nanocrystals (NCs). a-b**, $Mn^{2+}$-doped cores of wide-bandgap ZnSe are overcoated with narrower-gap CdSe shells of increasing thickness $h$. The conduction and valence band diagrams depict the notional electron and hole wavefunctions in these NCs. Room-temperature photoluminescence (PL) and absorption spectra (OD; optical density) from representative solutions of (**a**), $Mn^{2+}$-doped ZnSe cores only and (**b**), $Mn^{2+}$-doped cores with thick CdSe shells ($h \sim 8$ Å). With increasing $h$, the band-edge PL energy approaches and drops below the internal $Mn^{2+}$ $^4T_1 \rightarrow {}^6A_1$ emission at ~2.15 eV. **c**, Dependence of the band-edge exciton PL energy on the 1S absorption peak energy. Red points show NCs with $Mn^{2+}$-doped cores, and blue points show NCs with nonmagnetic 'reference' cores. The $Mn^{2+}$ emission at ~2.15 eV is shown by black squares.

**Figure 2. Magnetic-field and temperature-dependent Zeeman spin splitting, *ΔE_Z*, and magnetic circular dichroism (MCD) spectra from both magnetic and nonmagnetic core-shell NCs.** Plots show $\Delta E_Z$ versus $H$ at the fundamental 1S absorption peak from ZnSe/CdSe NCs at T = 3K (circles), 10 K (squares), 20 K (triangles), and 50 K (diamonds). The insets show absorption and MCD spectra at T = 3 K, for H = 2, 4, 6 T. The four panels show measurements on NCs having: **a**, undoped ZnSe cores and thin CdSe shells ($h<\sim1$ Å); **b,** undoped cores with thick shells ($h\sim7$ Å); **c,** $Mn^{2+}$-doped cores with thin shells; and **d,** $Mn^{2+}$-doped cores with thick shells. Note that the MCD and $\Delta E_Z$ inverts sign in $Mn^{2+}$-doped NCs when shell thickness increases (**c,d**). The black dashed lines in (**c,d**) show modified Brillouin-function fits to $\Delta E_Z$ at T=3K, using effective temperatures T + $T_0$ = 4 K and 9 K, respectively (see Methods).

**Figure 3. Field-dependent Zeeman splitting of the 1S absorption peak, *ΔE_Z*, and corresponding effective exciton *g*-factors at T=1.6 K. a**, The measured $\Delta E_Z$ for one series of $Mn^{2+}$-doped ZnSe/CdSe NCs having different CdSe shell thickness $h$ ranging from $h < 1$ Å (red) to $h \sim 8$ Å (purple). The black points show $\Delta E_Z$ for nonmagnetic 'reference' core-shell NCs. **b**, Effective exciton g-factors, $g_{eff}$, derived from the low-field Zeeman splitting at 1.6 K, plotted as a function of the 1S absorption edge energy. Also shown are $g_{eff}$ from the nonmagnetic 'reference' core/shell NCs (black points). $g_{eff}$ inverts when $h\sim2$Å.

**Figure 4. Energy level diagrams illustrating the Zeeman splitting at the NC absorption edge, in both an exciton and an electron-hole picture.** **a**, The five exciton levels of the 1S band-edge exciton in NCs. The $1^L$ and $1^U$ exciton states are primarily responsible for PL and absorption, respectively, and their separation gives the 'global' Stokes shift. MCD derives from the Zeeman splitting of the absorbing $\pm 1^U$ states. According to the experimental data, the order of the $+1^U$ and $-1^U$ states changes upon introduction of $Mn^{2+}$ ions into NCs with thin shells, but remains unchanged for the case of NCs with thick shells. In both cases, though, the Zeeman splitting in doped NCs greatly exceeds that in undoped NCs. **b**, Representation of the Zeeman splitting of the $1^U$ excitonic state in terms of the individual splittings of the $s_z = \pm\frac{1}{2}$ electron and $j_z = \pm\frac{3}{2}$ hole states that comprise the $1^U$ exciton, for the case of i) nonmagnetic NCs, ii) $Mn^{2+}$-doped NC with thin CdSe shells, and iii) $Mn^{2+}$-doped NC with thick CdSe shells. Independent of shell thickness, introduction of $Mn^{2+}$ ions into NCs changes the order of the +1/2 and -1/2 electron states, assuming that the sign of $\alpha$ is inverted in NCs compared to the bulk. For thin shells, however, the 'excitonic' exchange is still dominated by hole-$Mn^{2+}$ interactions; therefore, the resulting Zeeman splitting observed in MCD is qualitatively similar to that in bulk DMS materials. Alternatively, for thick shells the strength of the $Mn^{2+}$-hole interaction is decreased because holes become primarily shell-localized. As a result, the measured excitonic splitting is dominated by the 'sign-inverted' electron-$Mn^{2+}$ interaction. **c**, A diagram depicting the relative energies of quantum-confined electron and hole levels in the NC ($E_e$ and $E_h$) and the occupied and unoccupied Mn 3$d$ levels ($\varepsilon^+$ and $\varepsilon$). Red arrows show the virtual transitions that enter into calculations of the confinement-induced contribution of kinetic-exchange to $\alpha$ (described in text).

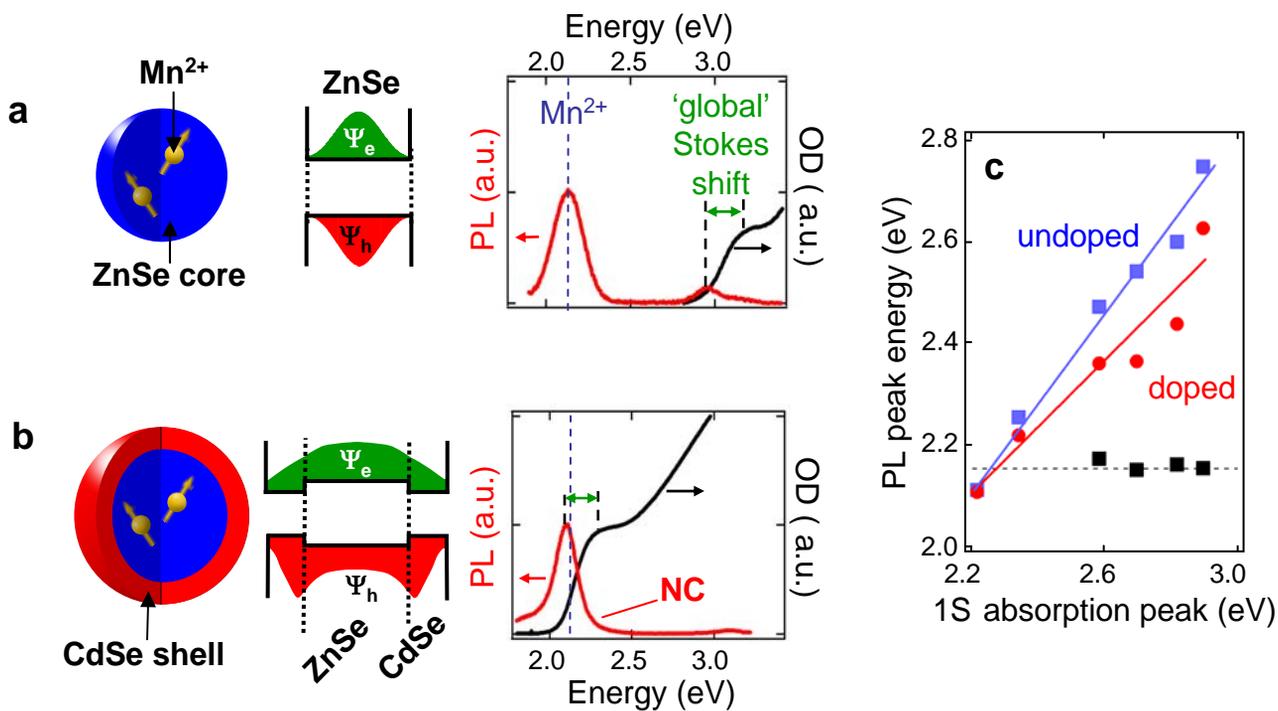

Figure 1

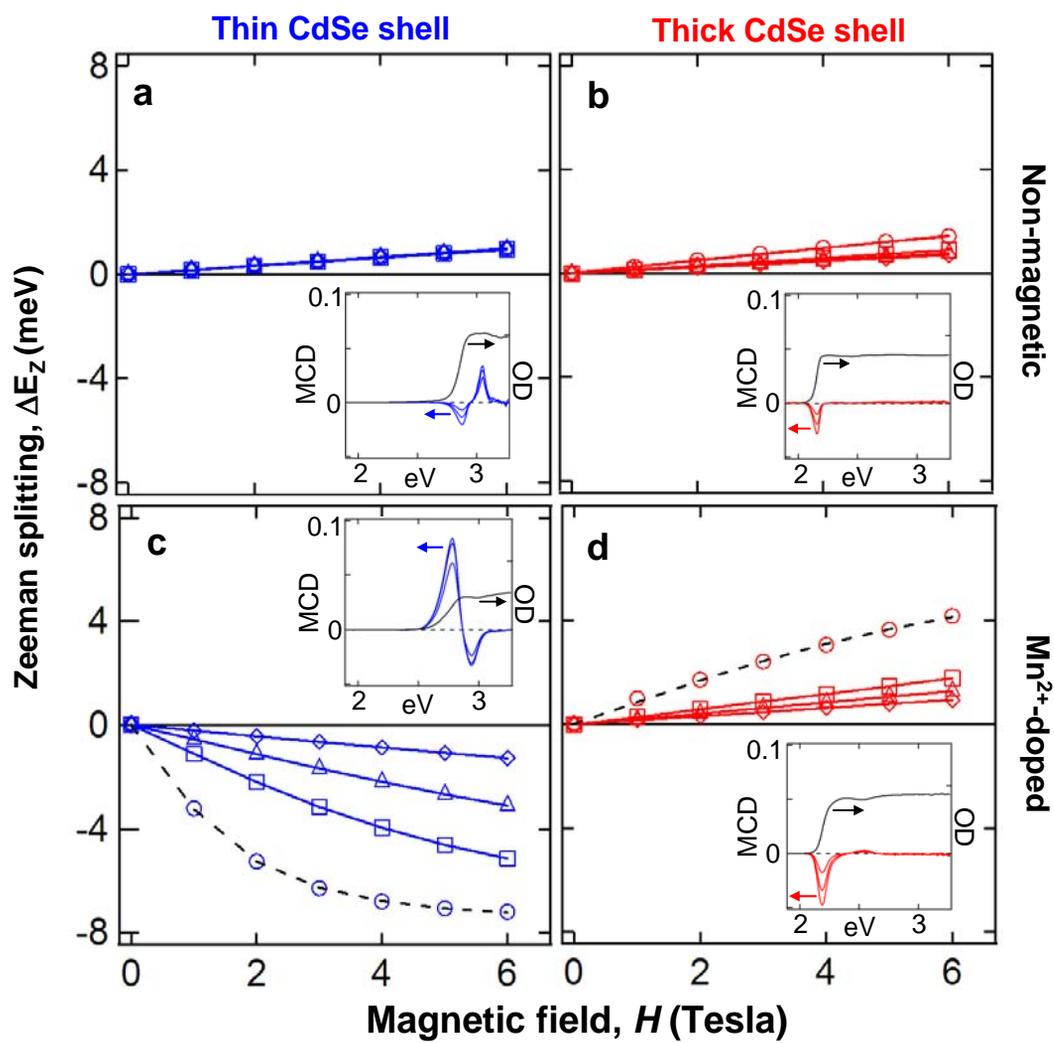

Figure 2

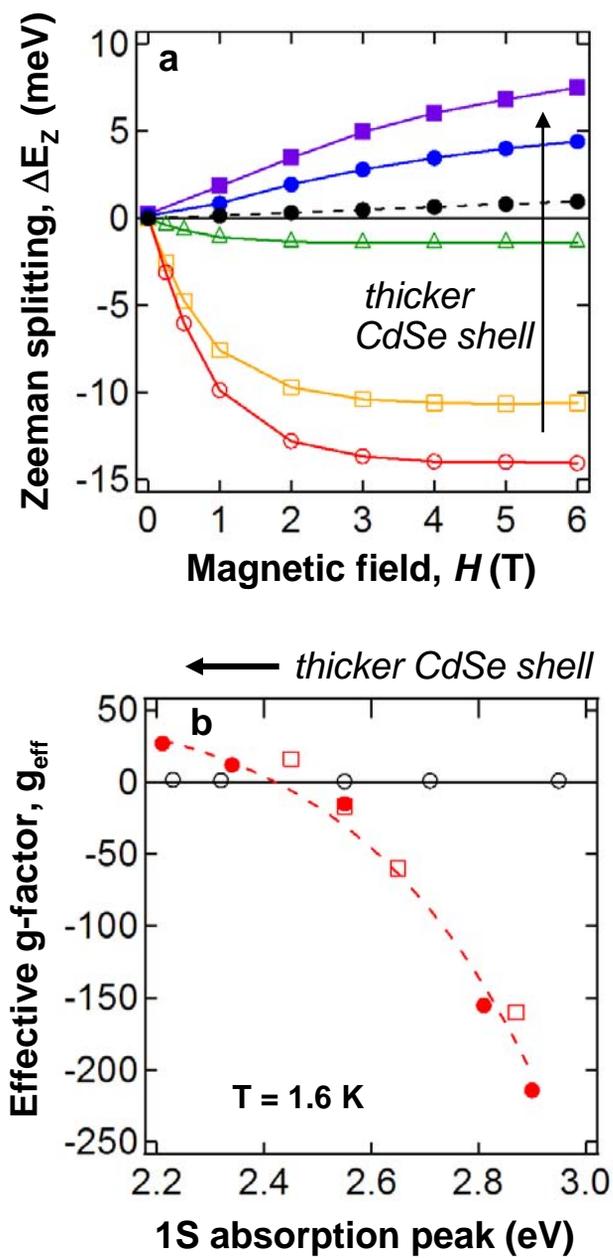

Figure 3

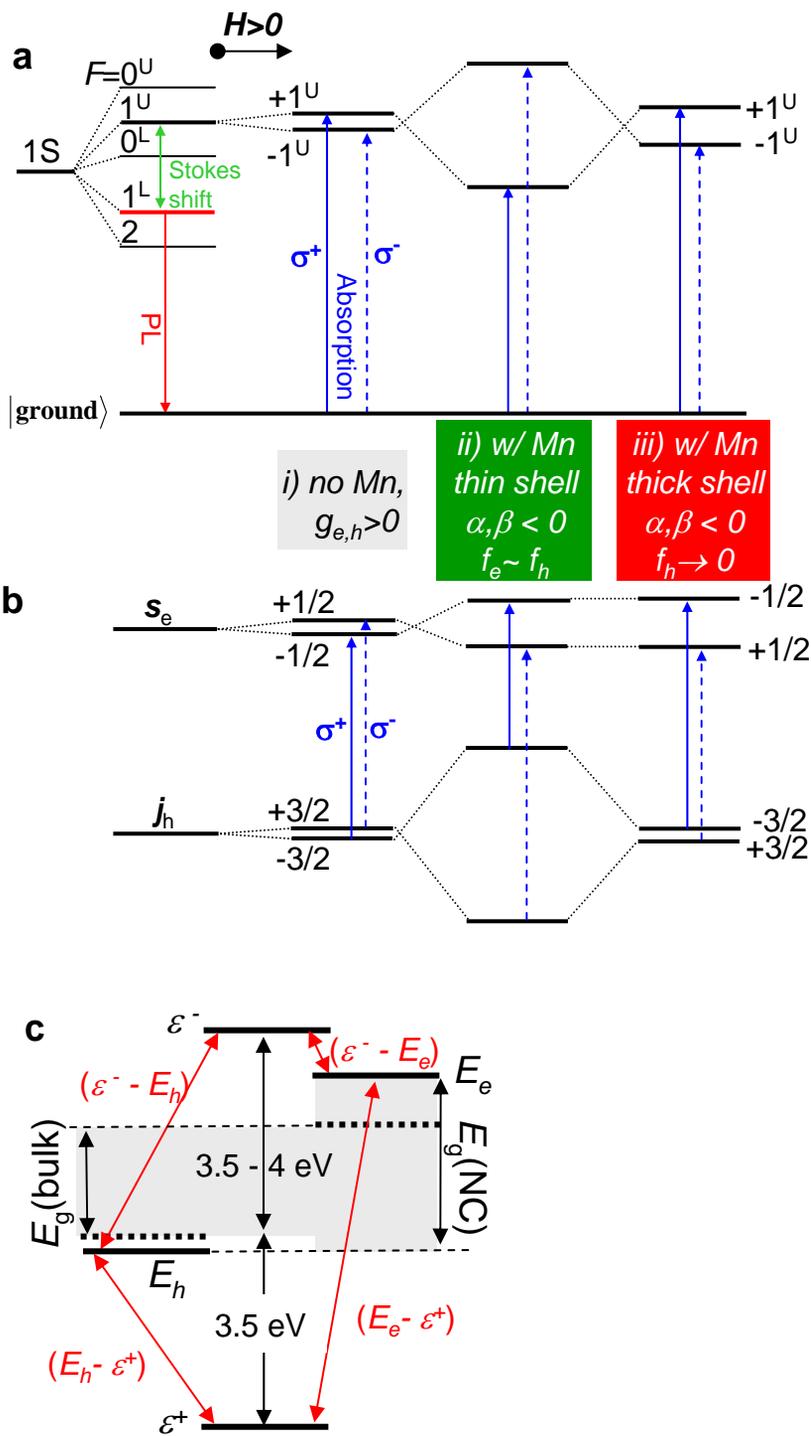

Figure 4

# SUPPLEMENTARY INFORMATION:

"**Tunable magnetic exchange interactions in manganese-doped inverted core/shell ZnSe/CdSe nanocrystals**", by D. A. Bussian et al.

**Electron paramagnetic resonance (EPR)** studies help to identify the crystalline environment of the $Mn^{2+}$ ions within these ZnSe/CdSe core/shell NC (see Fig. S1). The hyperfine interaction of unpaired *d*-electrons with the $^{55}Mn$ nuclear spin results in the well-known six-peak EPR spectra. The inter-peak spacing is sensitive to the crystalline environment of the $Mn^{2+}$ ions and can be used to help infer dopant location[1]. The 9.72 GHz EPR spectrum of the $Mn^{2+}$:ZnSe cores and $Mn^{2+}$:ZnSe/CdSe core/shell NCs are shown in Fig. S1 (black traces), along with a simple corresponding model (red traces). The spectra contain contributions from $Mn^{2+}$ in two distinct crystalline environments having hyperfine coupling constants of $61.4 \times 10^{-4}$ $cm^{-1}$ and $82.7 \times 10^{-4}$ $cm^{-1}$ in the ratio of ~80%:20% based on integrated signal strength. Although the exact magnitude of the $Mn^{2+}$ lines is not accurately reproduced (this particular simulation does not include the forbidden and the outer transitions, which strongly depend on the magnitude and anisotropy of the zero-field splitting), the overall position of the spectral patterns agrees well with the data.

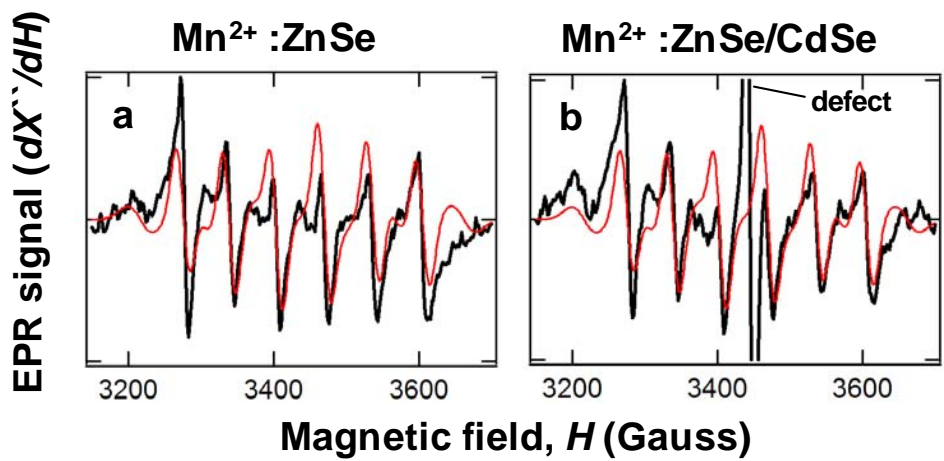

**Figure S1.** 9.72 GHz EPR spectra (black) and simulations (red) of **a**, $Mn^{2+}$:ZnSe cores (*h*=0) and **b**, $Mn^{2+}$:ZnSe/CdSe core/shell NCs (*h*~8 Å). The sharp peak at 3440 G (g = 2) in **b** is only observed in core/shell NCs and does not correlate with changes in the $Mn^{2+}$ signal. We tentatively attribute this feature to defects at the core/shell interface.

The crystalline environment 'seen' by the majority (~80%) of the $Mn^{2+}$ is characterized by the smaller hyperfine coupling constant (~62 x $10^{-4}$ $cm^{-1}$) and is associated with ions with distorted tetragonal symmetry as observed[2] in bulk ZnMnSe. Similar behavior is observed in all measured NCs (both core-only and core/shell). The larger hyperfine coupling constant (82.7 x $10^{-4}$ $cm^{-1}$) exhibited by ~20 % of $Mn^{2+}$ ions is consistent with a distorted octahedral symmetry[3], suggesting ions that lie near or at the NC surface. Given that these samples were thoroughly ligand-exchanged with pyridine and TOPO, it is unlikely that there is any significant contribution from surface-bound $Mn^{2+}$ and hence most $Mn^{2+}$ ions are located within the ZnSe core for all of our core-only and core-shell nanocrystals.

As for the "sharp defect" peak centered at g = 2, we do not have an exact explanation for the origin of this signal, other than this feature is always observed (with varying magnitude) in all the ZnSe/CdSe nanoparticles. Analysis of a large number of EPR spectra taken in different conditions and on different samples indicate that the magnitude of this single transition does not seem to be correlated with any changes of the $Mn^{2+}$ dopant EPR signal. This signal may be related to defects or electron trapping at the interface between the core and the outer shell of the nanocrystal, since no such signals are observed in simple ZnSe particles. One other possibility is that this signal is due to singly ionized selenium vacancy centers, as observed in ZnSe grown by molecular beam epitaxy[4].